\documentclass[a4paper,11pt]{article}
\usepackage{jheppub} 
\usepackage{lineno}
\usepackage{graphicx}
\usepackage{dcolumn}
\usepackage{bm}
\usepackage[dvips]{color}
\usepackage{cases}
\usepackage{stmaryrd}
\usepackage{mathrsfs}
\usepackage{amssymb}
\usepackage{amsmath}
\usepackage{txfonts}
\usepackage{slashed}
\usepackage{graphics}
\usepackage{bbold}
\usepackage{dsfont}
\linenumbers
\def\be{\begin{equation}}
\def\ee{\end{equation}}
\def\bfi{\begin{figure}}
	\def\efi{\end{figure}}
\def\bec{\begin{center}}
	\def\enc{\end{center}}

\def\af{\alpha}
\def\bt{\beta}
\def\ga{\gamma}

\def\tht{\theta}
\def\si{\sigma}

\def\ep{\epsilon}
\def\lam{\lambda}

\def\h{{1\over 2}}

\def\Or{\mathcal {O}}

\title{\boldmath Innovative Polarimetry for High‐energy Cosmic $\ga$ and $e^{+}/e^{-}$ Induced by Vector Photo‐productionn}

\author[a]{Dart-yin A. Soh\note{Corresponding author.}}
\author[b]{Zhaoyi Qu}
\affiliation[a]{Institute of Physics, Academia Sinica, Taiwan}
\affiliation[b]{Shandong Institute of Advanced Technology, Jinan, P. R. China}

\emailAdd{dartyinsoh@icloud.com}

\abstract{In this paper, we explore the possibility of measuring the complete polarizations of cosmic photons $\ga$ and the polarizations of cosmic electrons $e^{-}$ and positrons $e^{+}$. Our innovative Vector Meson Photo-production induced polarimetry enables people to measure the circular polarization component of a $GeV$ $\ga$ and to improve its linear polarization measurement, and thus enables people to measure the polarization of $GeV$ $e^{+}/e^{-}$ for the first time. We calculate the production process of $\pi^{+}\pi^{-}$ by a generally polarized photon near nucleon's field in a generalized \textit{VPD-SDMEs Factorization} with the fitted experimental data, so that it's partially model-independent. We also propose the observables and approach to measure their polarizations based on our calculations. Our new polarimetry of high-energy cosmic $\ga,e^{+},e^{-}$ will open a new window to reveal the mysteries and solve the puzzles of BSM new physics in particle physics and cosmology.}

\begin{document}
\nolinenumbers
\maketitle
\flushbottom

   \section{Introduction}
	\textbf{Motivations} Polarization measurements of cosmic photons had achieved and will continue to achieve a lot of important discoveries and progress in astrophysics and cosmology over the past few decades, including the E and B modes of CMB anisotropy to establish the standard model of cosmology and to test the inflation models \cite{PolmetryRev1}. Polarization provides more complete and cutting-edge probe of new physics beyond spectroscopy, photometry and imaging/mapping measurements, in the current multi-messenger and multi-variable analysis era. The astrophysical sources like pulsars, AGNs, GRBs can produce high-energy photons with (circular) polarizations, and accelerate polarized $e^{+}/e^{-}$, and the measurements can help us to solve the remained problems in astrophysics. New physics of beyond Standard Model of Particle Physics (BSM) processes like dark matter annihilation and axion-like particle Primakoff process can produce circular polarized photons and polarized $e^{+}/e^{-}$, the detection can explore new physics of BSM and cosmology. \cite{circpolphys1,circpolphys2,KWNg} For the \textit{spin polarization} of the cosmic electrons $e^{+}$/positrons $e^{-}$, there's no detectable process sensitive to the spin-polarization states of the high-energy $e^{+}/e^{-}$. Therefore, very few works on the literatures discuss the production mechanism of polarized cosmic $e^{+}/e^{-}$.  Astrophysical sources and new physics productions of the polarized high-energy $\ga/e^{+}/e^{-}$.
	
	\textbf{Experimental New Physics Hints} Recently the Alpha Magnetic Spectrometer (AMS-02) experiment on international Space Station (ISS) had measured spectra of high-energy electrons $e^-$ and positions $e^{+}$ from space, and had detected plenty $\Or(100)MeV$ to $\Or(10)GeV$ photons from space using $\gamma\rightarrow e^{+}e^{-}$ conversion process  in the materials of the detector. The polarization measurement of the photons sources is progressing in spite of difficulty caused by the small open angle between the $e^{+}e^{-}$ products converted by the photons. 
	On the other hand, it's suggested from the DR2 data that the related low-energy component ($<20GeV$) and the related high-energy component ($>20GeV$) have quite different energy distributions. The $e^{+}/e^{-}$ from these 2 components have likely different polarizations rather than are both unpolarized. However, the polarization states from $\mathscr{O}(10)GeV$ $e^{+}/e^{-}$ from space have never been measured. It's important to perform this measurement to detect potential new physics of dark matter or astrophysical processes. On a particle detector in space like AMS, we can use the polarization of the bremsstrahlung photons $\gamma$ emitted by $e^{+}/e^{-}$ source to detect the its polarization state. However, to distinguish a polarized $e^{+}/e^{-}$ from the unpolarized case, the circular polarization of its bremsstrahlung $\gamma$ has to be measured.
	
	\textbf{Polarimetry Review} The optical polarimetry, such as the photoelectric effect, Thomson scattering, Compton scattering, can only cover the energy range from microwave up to a few $MeV$. $e^{+}e^{-}$ pair production polarimetry in the electro-magnetic (EM) field of a nucleus can measure the linear polarization of a photon in the medium- and high-energy region ($\sim 1MeV$ to $1 GeV$) but suffers from the low efficiency at $\mathscr{O}(100)MeV$ due to the small open angle between the $e^{+}e^{-}$ pair and their \textit{multi-scatterings} for  high-energy photons. Recently people propose a few improvements of resolution using TPC and other techniques like AdEPT and MSD in $e^{+}e^{-}$ conversions. However, at high-energy region $0.5GeV\sim \mathscr{O}(10) GeV$ the polarimetry is lacked, and furthermore, there is still no full polarimetry including circular polarization in the medium- and high-energy region. It's well-known that only the linear (plane) polarization components of the photons can be measured by $e^{+}e^{-}$ conversion. The circular polarization component of the photons is not accessible by $e^{+}e^{-}$ conversion \cite{linee}. 
	On the other hand, we still lack the polarimetry proposed for high-energy cosmic $e^{+}/e^{-}$. There are 3 techniques to measure the polarization of electron beams based on scattering the electrons from: unpolarized nuclei (Mott scattering), polarized atomic electrons (M\o{}ller scattering), and polarized photons (Compton scattering). Mott polarimetry is effective only at low electron energy ($,\mathscr{O}(10) MeV$) and for transverse polarization, while M\o{}ller and Compton polarimetry are effective for high-energy longitudinally polarized electrons and precise from QED calculations. However, they require the polarized electron targets or the polarized lasers which are impossible for the high-energy cosmic $e^{+}/e^{-}$ detection.
	In this paper we will propose our innovative polarimetry for the circular polarization of cosmic $\ga$ and polarization of cosmic $e^{+}/e^{-}$ that will open a new window for the astroparticle study! We will discuss the physics of productions of circularly polarized photons and polarized $e^{+},e^{-}$ sources in our paper \cite{polphys}.
	
	It's overlooked that a high-energy photon can produce a pair of pions $\pi^{+}\pi^{-}$ via the strong interaction when colliding with a nucleus and its polarization can thus be probed. This process is dominated by the vector-meson $\rho^0$-photoproduction; and similarly photons can also produce $\pi^{0}\pi^{+}\pi^{-}$ dominated by $\omega$-photoproduction and $K^{+}K^{-}$ dominated by $\Phi$-photoproduction. This idea is referred to as \textit{$V$-Photoproduction induced Photon Polarimetry}. We will focus on the $\pi^{+}\pi^{-}$ production here, however, considering the complexity in the $\pi^{0}\pi^{+}\pi^{-}$ analysis and the small cross section in the $K^{+}K^{-}$ production. Due to the heavy mass of the vector-meson ($m_{\rho}=770MeV$), the open angle between $\pi^{+}\pi^{-}$ pair is expected much larger than that in the $e^{+}e^{-}$ conversion, and result in much better angular resolution to measure the $\gamma$ polarization. Moreover, the circular polarization component of a photon can transmit to the 3-spin state of a massive vector-meson, hence we can measure the circular polarization of a photon using the angular distribution of the pion pair $\pi^{+}\pi^{-}$ conversion. We will calculate event rate of the process for our polarimetry in this article.
	
\begin{figure}[h]
		\centering
		\includegraphics{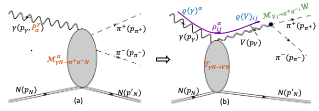}
		\caption{Diagrams of the process of polarized photon conversion to $\pi^{+}\pi^{-}$. (a) The whole process $\gamma N \rightarrow \pi^{+}\pi^{-} N$ with the photon polarization vector $P_{\af}^{\ga}$; (b) The $V$ photoproduction dominates the process. The diagram illustrates the VPD-SDMEs Fractorization, as the spin density matrices of photon ($\varrho(\ga)$) and of $\rho^{0}$ ($\varrho(V)$) transferred between production and decay amplitudes.}
		\label{fig:diagVPD}
	\end{figure}

	The photoproduction of a meson pair in the nucleonic field can be calculated either in perturbation models like chiral perturbation theory and the Nambu-Jona-Lasinio model, or in Regge Theory non-perturbatively. There's no first-principle calculation like Lattice QCD in this process. The cross sections peak at low momentum transfer $|t|$ region, where the strong interaction enters the strong coupling limit and perturbation method tends to break down. 	
	Therefore, there are large uncertainties in the phenomenological strong-interaction models of the photoproduction and its decay. To reduce the theoretical uncertainty, and to calculate the differential cross section with the polarised photon, in this article we adopt the factorisation method to separate the production and decay in the calculations of the cross sections, in which the intermediate vector meson $\rho^{0}$ state dominate production of the $\pi^{+}\pi^{-}$ pair, as illustrated with the diagrams in Fig.\ref{fig:diagVPD}. And then we calculate the production cross section and the spin density matrix elements (SDMEs) by parameterization with the available experimental data fittings, for this process. 
	We then implement the calculations on Geant4 simulation generator to enable us to do the measurement of the photon polarization and the $e^{+}/ e^{-}$ polarization on AMS detector. 
	
	\section{$V$-Photoproduction Induced Polarimetry of $\ga$ and $e^{+}/e^{-}$}
	
	\subsection{Polarization Formulism}
	It was well understood that the \textit{spin} state is a property of an individual particle, while the \textit{polarization} state is a statistical property of particles in a production source, \textit{i.e.} of a beam. In the case of an individual particle, the polarization is defined as a time average over many observations of its spin. Thus, it is possible for an individual particle to be unpolarized \cite{polbeam}. For an individual particle (in a typical 2-level systam like $\ga/e^{+}/e^{-}$) with specific spin state in an arbitrary bases $\zeta_{1},\zeta_{2}$ (either left/right $L/R$ or plain wave $x/y$ for $\ga$), \textit{i.e.} 
	\begin{equation}
	|\zeta\rangle=\af|\zeta_{1}\rangle+\bt|\zeta_{2}\rangle,
	\end{equation} 
	we can define the density operator for its polarization state 
	\be 
	\varrho=|\zeta\rangle \langle\zeta | = \h (P_{0}\textbf{1}+ P_{i}\sigma_{i})=\h
	\left( \begin{array}{cc}
		P_{0}+P_{3} & P_{1}-\mathrm{i} P_{2}\\
		P_{1}+\mathrm{i} P_{2} & P_{0}-P_{3}
	\end{array}\right)
	\ee 
	where $\sigma_{i},i=1,2,3$ are the Pauli matrices. This is the case for \textit{pure polarization} with $|\vec{P}|=1$, and $P_{0}=2(|\af|^2+|\bt|^2)$, $P_{3}=2(|\af|^2-|\bt|^2)$, $P_{2}=2\Im(\af^{*}\bt)$, $P_{1}=2\Re(\af^{*}\bt)$ \\
	The generalization is \textit{polarization} with $|\vec{P}|<1$ for statistical mixed state as mixture of pure states,
	since any statistical mixture can be obtained by incoherent quantum superposition of two appropriately chosen pure states $\varrho=w_{1}\varrho_{1}+w_{2}\varrho_{2}$, $\vec{P}=w_{1}\vec{P}_{1}+w_{2}\vec{P}_{2}$ with $|\vec{P}_{1}|=|\vec{P}_{2}|=1$. The polarization vector $\vec{P}$ with normalization $Tr[\varrho]=P^{0}=1$ is called \textit{Stokes vector}. The unpolarized particle has a unity density state $\varrho=\textbf{1},\vec{P}=0$. \\
	The polarized cross section of a process with polarized particles is calculated by tracing the helicity amplitude square with this polarization density matrix. The polarized cross section with the polarization density $\varrho$ of an initial state particle and helicity amplitude $\mathscr{M}^{i}$ ($M^{ij}=\mathscr{M}^{i}\mathscr{M}^{*j}$) is then
	\be
	d\si=Tr[\varrho M]\equiv \mathscr{M}\varrho \mathscr{M}^{\dagger}=\h [P_{0}(M^{11}+M^{22})+P_{3}(M^{11}-M^{22})]+P_{1}(M^{12}+M^{21})+i P_{2}(M^{21}+M^{12})
	\ee
	Following this formalism, the polarization density of the final state particle(\textit{e.g.,} $V$ meson) is then transformed from the polarization of the initial state (\textit{e.g.,} $\ga$), and can be calculated as 
	\be \label{rhof}
	\varrho(f)=\mathscr{T}_{norm}\varrho(i) \mathscr{T}^{\dagger}_{norm} \Rightarrow \varrho(V)=\mathscr{T}_{norm}\varrho(\ga) \mathscr{T}^{\dagger}_{norm}
	\ee
	where $\mathscr{T}$ ($\mathscr{T}_{norm}$) is the (normalized) production amplitude with final state helicity indices. After $\varrho(\ga)$ is decomposed into the Stokes vector of photon $\vec{P}^{\ga}$, the spin density matrix elements are defined as:
	\be
	\rho^{\af}_{ij}=\mathscr{T}_{i}(\h\textbf{1},\h\si^{\af}) \mathscr{T}^{\dagger}_{j},  \qquad \varrho(V)_{ij}=\rho^{0}_{ij}+\sum_{\af=1}^{3} P_{\af}^{\ga}\cdot\rho^{\af}_{ij}
	\ee
	
	\subsection{Polarization of $e^{+}/e^{-}$ via Bremsstrahlung \& Discussion of $e^{+}e^{-}$ Conversion}
	The high-energy cosmic electrons/positrons emit polarized bremsstrahlung photons in the EM field near nuclei in the detectors.  $e^{+}/e^{-}$ has $\pm$ helicities, its polarization vector $\vec{P}^{e}$ is thus defined in this spin basis and its physical meaning is as: $P_{1}^{e}$ is the \textit{longitudinal} polarization component along $e^{+}/e^{-}$ momentum while $P_{2}^{e}$ and $P_{3}^{e}$ are the \textit{transverse} polarization components in and perpendicular to the interaction plane. It's indicated that the \textit{spin-polarization} states $P_{i}^{e}$ of $e^{+}/e^{-}$ can transform via the bremsstrahlung interaction matrix $T$ to the polarization $\vec{P}^{\ga}$ of the bremsstrahlung photon when it's in the $L/R$ basis \cite{polbrem1,polbrem2,polbrem3}:
	\be \label{bremdyn}
	\begin{pmatrix}
		P_{0}^{\ga}\\
		P_{1}^{\ga}\\
		P_{2}^{\ga}\\
		P_{3}^{\ga}
	\end{pmatrix} 
	= T_{S} d E_{\ga} d t
	\begin{pmatrix}
		T_{I} & 0 & 0 & 0\\
		T_{D} & 0 & 0 & 0\\
		0 & 0 & 0 & 0\\
		0 & -T_{L} & -T_{T} & 0
	\end{pmatrix} 
	\begin{pmatrix}
		P_{0}^{e}\\
		P_{1}^{e}\\
		P_{2}^{e}\\
		P_{3}^{e}
	\end{pmatrix} 
	\ee
	where $T_{I},T_{D},T_{L},T_{T}$ are the matrix elements and $T_{S}$ is the factor, which are determined by the interaction dynamics and depend on various kinematic variables. They were calculated earlier in the Born approximation, and later were updated when the effects of anomalous magnetic moment and 2 form-factors of the proton target, and also the electron target field are considered \cite{newbrem1,newbrem2,newbrem3}. However, the form of the interaction matrix $T$ is determined by the spin symmetry in bremsstrahlung process: both unpolarized and polarized $e^{+}/e^{-}$ can generate \textit{linear} polarization component $P_{1}^{\ga}$ of bremsstrahlung $\ga$ via $T_{D}$ from $P_{0}^{e}$; while both longitudinal and transverse components $P_{1}^{e}$ and $P_{2}^{e}$ can generate \textit{circular} polarization component $P_{3}^{\ga}$ of $\ga$ via $T_{L}$ and $T_{T}$. The perpendicular transverse component $P_{3}^{e}$ is silent and the $\pi/4$ linear component $P_{2}^{\ga}$ of bremsstrahlung is absent. The degrees and dependence of the circular and linear polarizations of bremsstrahlung were also confirmed by measurements \cite{bremexp1,bremexp2}. 
	
	Therefore, in a high-energy detector detecting the $GeV$ cosmic electrons/positrons, their spin-polarization states which reveal new physics in their production processes, can be measured and only can be measured by the polarization states of their bremsstrahlung photon emissions effectively. However, linear polarization component of the bremsstrahlung $\ga$ which $e^{+}e^{-}$ conversion is accessible, is incapable even to differentiate whether the cosmic electrons/positrons are polarized. The circular polarization component measurement of the bremsstrahlung in our proposed $V$-photoproduction induced photon polarimetry is thus necessary in the cosmic $e^{+}/e^{-}$ polarimetry! They form a whole innovative polarimetry of $GeV$ cosmic $\ga/e^{+}/e^{-}$. 
	
	\textbf{Discussion of $e^{+}e^{-}$ Conversion: only Linear Polarization Accessible.}
	The photon $e^{+}e^{-}$ conversion in the field near a nucleus is the inverse process of the bremsstrahlung process, so the interaction matrix for this process is just the inverse of the bremsstrahlung matrix $T$. The case that the spins of $e^{+}e^{-}$ product are summed over corresponds to the bremsstrahlung of unpolarized $e^{+}/e^{-}$. It's obvious from this observation that only linear component $P_{1}^{\ga}$ of the photon polarization accesses the angular distributions of unpolarized $e^{+}e^{-}$ product, and the circular component $P_{3}^{\ga}$ could be accessed when the polarization of the created $e^{+}$ or $e^{-}$ is measured, which was analysed in the triplet production \cite{econvpol}, but it's not applicable in the astrophysical detectors.  
	We also recalculated the helicity amplitude with the photon polarization vector $P^{\ga}_{i}$ and thus obtained the polarized differential cross section in term of $P^{\ga}_{i}$: $d\si(P^{\ga}_{i})$. We verified that $d\si(P^{\ga}_{i})$ is independent of the circular component $P^{\ga}_{3}$.
	Therefore, our polarimetry is the only method to measure the circular polarization of high-energy cosmic photons.
	
	\section{Differential Cross Section of the Polarized Process}
	
	\subsection{VPD-SDMEs Factorization}
	The $\pi^{+}\pi^{-}$ pair conversion of a photon near a nucleus is complicated since it involves the $\pi^{+}\pi^{-}$ photoproduction by the photon's interactions in the nucleus' EM field (the $t$ and $u$ channels of $\pi^{+}\pi^{-}$ production via a virtual photon) \cite{empipi} and  in the nucleus' nucleonic field, showed in (a) of Fig.\ref{fig:diagVPD}. However, the latter dominate via the photoproduction of an intermediate vector meson $V$ in the $s$ channels of $\pi^{+}\pi^{-}$, due to the symmetry nature of the vector mesons $V(\rho^{0}, \omega,\Phi)$ as hadronic photon. This is the essential physics of $V$ Photoproduction Dominance (VPD) in the photon $\pi^{+}\pi^{-}$ conversion, which is empirically tested \cite{expreview}. Please note that this should to be differentiated from the Vector Meson Dominance (VMD) model which can calculate the $V$ photoproduction via the $VN\rightarrow VN$ strong process, using the $\ga-V$ correspondence. We also ignore the tiny interference between $\rho^{0}$ and $\omega$.
	
	The polarization $\varrho(\ga)$ (equivalently $P^{\ga}$) of the photon $\ga$ transfers to the the polarization $\varrho(V)$ of the vector meson, which can be off-shell and has 3 polarization modes (with index $i=-1,0,1$) due to its mass. The \textit{massive} property of the "heavy photon" $V(\rho^{0})$ is essential to make the circular polarization component of $\ga$ accessible in its decay product $\pi^{+}\pi^{-}$ pair. 
	
	To simplify the calculation of our complicated process $\gamma N \rightarrow \pi^{+}\pi^{-} N$, in the observation of VPD, the whole process can be decomposed into the $V(\rho^{0})$ photoproduction and its subsequent $\pi^{+}\pi^{-}$ decay $\gamma(p_{\gamma}) N(p_{N}) \rightarrow [\rho^{0}(p_{V})\rightarrow \pi^{+}(p_{\pi^{+}})\pi^{-}(p_{\pi^{-}})] N(p'_{N})$.
    Hence, applying the generated \textit{Factorization Theorem} (FT) \cite{genFT} accounting the intermediate resonance $V$'s spin $i,j$, and \textit{Breit-Wigner}(BW) \textit{propagator Approximation} with its mass $m_{V}$ and constant decay width $\Gamma_{V}$,we can partially \textit{factorize} the squared amplitude of the process into the production and decay amplitudes $\mathscr{T},\mathscr{M}$:
	\be \label{Msqcal}
	\mathscr{M}_{\ga N\!\rightarrow\pi^{+}\pi^{-} N}\cdot\varrho(\ga) \cdot \mathscr{M}^{\dagger}_{\ga N\!\rightarrow\pi^{+}\pi^{-} N}=\sum_{ij}\mathscr{T}_{\ga N\!\rightarrow V_{i} N}\cdot\varrho(\ga) \cdot\mathscr{T}^{\dagger}_{\ga N\!\rightarrow V_{j} N} \times\frac{\mathscr{M}_{V_{i}\!\rightarrow\pi^{+}\pi^{-}} \mathscr{M}_{V_{j}\!\rightarrow\pi^{+}\pi^{-}}^{\dagger}}{|p_{V}^{2}-m_{V}^{2}-\mathrm{i} m_{V}\Gamma_{V}|^2}
	\ee
	The factorization is not straightforward due to this the spin correlations, nevertheless, it can still achieve by means of the polarization density transfers! Regarding the definition of $\rho(V)$ in Eq.(\ref{rhof}), the helicity production amplitudes with spin correlations can be factorized into the unpolarized production cross section as the normalization factor, and the $V$ spin density matrix: $\mathscr{T}_{\ga N\!\rightarrow V_{i} N}\cdot\varrho(\ga) \cdot\mathscr{T}^{\dagger}_{\ga N\!\rightarrow V_{j} N}\varpropto d\si_{\ga N\!\rightarrow V N}^{unpol}\varrho(V)_{ij}$! The physics of our factorization is illustrated in the Diagram (b) of Fig.\ref{fig:diagVPD}.
	Then it's expected that only the unpolarized 2-to-2 cross section $d\si_{\ga N\!\rightarrow V N}^{unpol}$ present in the factorized 2-to-3 squared amplitude integrating 3-body phase space $d\Phi_{3}(p_{V},p_{\pi^{+}},p_{\pi^{-}})$ with phase space factor $K$:
	\be  \label{xseccal}
	  \begin{split}
	d\si^{2\to 3}&=K\mathscr{M}_{2\to 3}\cdot\varrho(\ga) \cdot \mathscr{M}^{\dagger}_{2\to 3}d\Phi_{3}(p_{V},p_{\pi^{+}},p_{\pi^{-}})\\
	&=d\si_{\ga N\!\rightarrow V N}^{unpol}\frac{\Gamma_{V}(\mu)d\mu^{2}}{(\mu^{2}-m_{V}^{2})^{2}+m_{V}^{2}\Gamma^{2}(\mu)}\frac{\sum_{ij}\varrho(V)_{ij}\mathscr{M}_{V_{i}\!\rightarrow\pi^{+}\pi^{-}} \mathscr{M}_{V_{j}\!\rightarrow\pi^{+}\pi^{-}}^{\dagger}}{\Gamma_{V}(\mu)}d\Phi_{2}(p_{V};p_{\pi^{+}},p_{\pi^{+-}})
	   \end{split}
	\ee	
	The calculations show the spin decorrelations of between the $V$ production amplitude and its decay amplitude, when the spin $i,j$ in the production dynamics are encoded in $\rho(V)$. The convolution of $\rho(V)$ and the decay amplitude had been analysed using the symmetry properties of the helicity amplitudes by Schilling \textit{et.al.}\cite{schilling}. Let $\theta,\phi$ denote the polar angle and azimuth angle of one pion in the $V$'s rest frame after the $p_{V}$ defined in production c.m. frame, so $\phi$ actually the angle between the decay plane and the production plane in this c.m. frame.
	They calculated the decay angular distribution function of $\pi^{+}\pi^{-}$ pair $W(\cos\theta,\phi)$ with respect to  $\varrho(V)=\mathscr{T}\cdot\varrho(\ga) \cdot\mathscr{T}^{\dagger}$ (decomposing to SDMEs), by Wigner rotation function:
	\be \label{calW} 
	\frac{1}{\Gamma_{decay}}\frac{d\Gamma_{decay}}{d\tht d\phi}=\frac{\mathscr{M}_{decay}\cdot\varrho(V)\cdot \mathscr{M}_{decay}^{\dagger}}{32\pi^{2}\Gamma_{decay}}\frac{|\vec{p}_{\pi}^{*}|}{\mu^{2}}\equiv W(\theta,\phi)=W^{0}(\theta,\phi)+\sum_{\af=1}^{3}P^{\ga}_{\af} W^{\af}(\theta,\phi)
	\ee
	where the polarized function $ W^{\af}(\theta,\phi)$ are calculated with polarization vector component $L/R$ basis $P^{\ga}_{\af}$:
	\be \label{Wfunc}
	\begin{array}{rl}
		W^{0}(\theta,\phi)=&\frac{3}{4\pi}[\h(\rho^{0}_{00}(3\cos^{2}\tht-1)+\sin^{2}\tht)-\sqrt{2}\Re\rho^{0}_{10}\sin2\tht\cos\phi-\rho^{0}_{1-1}\sin^{2}\tht\cos2\phi], \\
		W^{1}(\theta,\phi)=&\frac{3}{4\pi}(\rho^{1}_{00}\cos^{2}\tht+\rho^{1}_{11}\sin^{2}\tht-\sqrt{2}\rho^{1}_{10}\sin2\tht\cos\phi-\rho^{1}_{1-1}\sin^{2}\tht\cos2\phi), \\
		W^{2,3}(\theta,\phi)=&\frac{3}{4\pi}(\sqrt{2}\Im\rho^{2,3}_{10}\sin2\tht\sin\phi+\Im\rho^{2,3}_{1-1}\sin^{2}\tht\sin2\phi)
	\end{array}
	\ee
	$\rho^{\af}_{ij}$ are the SDMEs of $V$ photoproduction. 
	The symmetry property reduces the 36 SDMEs to only 11 independent ones $\rho^{0}_{00},\rho^{0}_{1-1},\Re\rho^{0}_{10},\rho^{1}_{00},\rho^{1}_{11}\rho^{1}_{1-1},\Re\rho^{1}_{10},\Im\rho^{2}_{10},\Im\rho^{2}_{1-1},\Im\rho^{3}_{10},\Im\rho^{3}_{1-1}$ entering Eq.(\ref{Wfunc}). An unpolarized photon produces angular distribution $W^{unpol}(\theta,\phi)=W^{0}(\theta,\phi)$, while ($L/R$) circularly polarized photon ($P^{\ga}\equiv P^{\ga}_{3}$) produces $W^{\pm circ}(\theta,\phi)=W^{0}(\theta,\phi)\pm P^{\ga}W^{3}(\theta,\phi)$. 
	
	Thus, finally the 5-dimensional differential cross section of process $\gamma N \rightarrow \pi^{+}\pi^{-} N$ is factorized as:
	\begin{equation} \label{sigcal}
	\frac{d\sigma[\varrho(\gamma)]}{d\mu d\Omega_{V}d\Omega_{\pi}}=(16\pi^{2}m_{V}\mu)\left[
	\frac{\Gamma(\mu)}{(\mu^{2}-m_{V}^{2})^{2}+m_{V}^{2}\Gamma^{2}(\mu)}\right]
	\left[\frac{dt}{d\Omega_{V}(t)} \cdot 
	\left. \frac{d\sigma^{UN}}{dt} \right\bracevert_{mea}
	\cdot P^{\ga}_{\alpha}(\gamma)\right] [W^{\alpha}(\Omega_{\pi},\rho^{\alpha}_{ij}[V])]
	\end{equation}
	The first term \textit{i.e.} the phase space integration factor is simply $Pf=16\pi^{2}m_{V}\mu$, the second term denoted as $\widehat{BWp}$ is Breit-Wigner propagator for $V$, and the last 2 terms are $V$ production and decay cross section respectively, where the Jacobian factor in Eq.(\ref{sigcal}) $J^{2\rightarrow 2}\equiv dt/d\Omega_{V}(t)=E_{V}^{cm}(\mu)P_{V}^{cm}(\mu)/\pi$ depends on the generated $\mu$.
	The $\ga N \rightarrow V N$ strong interaction production is still complicated:
	we use the experimental measured unpolarized production cross section $\frac{d\sigma^{UN}}{dt}$ to calculate $d\si_{\ga N\!\rightarrow V N}^{unpol}$, fitted with momentum transfer $t$ dependence rather than calculate in various models. We emphasize the $\pi^{+}\pi^{-}$ angular and SDMEs dependence in $W^{\alpha}$, and the SDMEs are critical to achieve our polarization measurement. Fortunately there are completed and ongoing experiments measuring the SDMEs for $\rho^{0},\omega$ and $\Phi$ mesons. We use state-of-the-art results to fit them or calculate them in Regge Theory. 
	
    The approach to calculate the differential cross sections of $\ga N \rightarrow \pi^{+}\pi^{-}N$ mentioned above in this section can be called as \textit{VPD-SDMEs Factorization}. It's straightforward extend our calculations of VPD-SDMEs Factorization in processes $\ga N\rightarrow [\omega\rightarrow\pi^{0}\pi^{+}\pi^{-}]N$ and $\ga N\rightarrow [\Phi\rightarrow K^{+}K^{-}]N$. For the former $\theta,\phi$ are defined using the normal of the $\pi^{0}\pi^{+}\pi^{-}$ plane in $V$'s rest frame and the final-state phase space integration in Eq.(\ref{xseccal}) is just simply replaced $d\Phi_{2}(p_{V};p_{\pi^{+}},p_{\pi^{+-}})\Rightarrow d\Phi_{3}(p_{V};p_{\pi^{0}},p_{\pi^{+}},p_{\pi^{+-}})$, while for the latter the kinematics and forms of calculations remain the same.
	
	\subsection{Parameterization Model of the $VN$ Production} \label{sssecXsec}
	In 1970s SLAC experiment performed a series of measurement of the photoproduction of $\rho^{0}\rightarrow \pi^{+}\pi^{-}$, $\omega\rightarrow \pi^{0}\pi^{+}\pi^{-}$ and $\Phi\rightarrow K^{+}K^{-}$ at the photon energy $E_{\ga}$ of $2.8GeV$, $4.7GeV$ and $9.3GeV$ \cite{slac}. These energy points can be well extrapolated to the energy range from $\Or(100MeV)$ to $\Or(10GeV)$ for the cosmic photons or bremsstrahlung photons. We present a 2-dim $E_{\ga}-t$ fitting for measured cross sections $ \frac{d\sigma^{UN}}{dt}$ and also the 11 SDMEs.
	
	SLAC experiment use the proton target collisions. It was found that the differential cross section $\frac{d\sigma^{UN}}{dtdM_{\pi^{+}\pi^{-}}}$ of momentum transfer $t$ \& dipion mass $M_{\pi^{+}\pi^{-}}$ is well presented by the exponential decreasing form $\left. \frac{d\sigma^{UN}}{dtdM_{\pi^{+}\pi^{-}}}\right\bracevert_{t=0}e^{-\kappa|t|}$, which is well-known a feature of simple 2-to-2 high-energy scatterings\cite{uexp}. There are 4 methods to extract the limit value $\left. \frac{d\sigma^{UN}}{dt}\right\bracevert_{t=0}$ and the slope $\kappa$ for the $t$ differential cross section from the experimental data. We use the fitting results of Parameterization method, in which they fitted the Dalitz plot to a matrix element consisting of phase space, and it's the only method independent of $(m_{V},\Gamma_{V})$ and Drell background. It was pointed out \cite{slac} that $\left. \frac{d\sigma^{UN}}{dt}\right\bracevert_{t=0}$ decrease and $\kappa$ increase along $E_{\ga}$ simply, so linear fitting with Orthogonal-Distance-Regression (ODR) to take the measurement uncertainties into account is adapted:
	\be
	\left. d\sigma^{UN}/dt\right\bracevert_{t=0}(E_{\ga})=149.594807(\pm 12.6865689)~ - ~ 6.1009918 (\pm 1.85449934)\cdot E_{\ga} \quad(\mu b/GeV)
	\ee
	\be
	\kappa(E_{\ga})=6.52319561977(\pm 0.398815734127) ~+~ 0.0883205413808 (\pm 0.0560001228847)\cdot E_{\ga} \quad(GeV^{-1})
	\ee
	Then the $\ga N \rightarrow V N$ production cross section is 
	\be \label{dsi2} 
	\frac{d\sigma^{UN}}{dt} [E_{\ga}] =\left. \frac{d\sigma^{UN}}{dt}\right\bracevert_{t=0}(E_{\ga})\cdot\exp[-\kappa(E_{\ga})|t|]
	\ee	
	To extend our calculation to a general photon energy $E_{\ga}$, we need to consider the case that only part of the Breit-Wigner propagator of intermediate $V$ can be covered at low $E_{\ga}$, and calculate a cross section correction factor $\widetilde{c\si}(E_{\ga})$ due to the reduction of available 5D phase space, by the $\mu$ integrations of the BW propagator bounded by $\mu_{lim}(E_{\ga})=m_{N}(\sqrt{(2E_{\ga}+m_{N})/m_{N}}-1)$:
	\be
	\widetilde{c\si}(E_{\ga})= \frac{ \tan^{-1}\left[\frac{2m_{N}(E_{\ga}+m_{N})-2m_{N}\sqrt{m_{N}(2E_{\ga}+m_{N})}-m_{V}^{2}}{m_{N}\Gamma_{V}}\right]+\tan^{-1}\left[4-4\frac{\Gamma_{V}}{m_{V}}\right] }{\frac{\pi}{2}+\tan^{-1}\left[4-4\frac{\Gamma_{V}}{m_{V}}\right] }
	\ee
	This correction is smooth: when $E_{\ga}$ large enough it restores to 1 and when $E_{\ga}$ is smaller then the lower threshold it restores to 0, it's thus physical.
	For the heavier nuclei target with atom number $A$, people recently studied the transparency ratio $R=\frac{\si_{\ga A\rightarrow VX}}{A \si_{\ga p\rightarrow VX}}$ \cite{genNA1,genNA2}. It was found that there was negligible $(t,E_{\ga})$ dependence in the transparency ratio $R$. From their data we perform a fitting with optimized fitting function
	\be
	R(A)=a+bA+c/(A+A_{0}), \quad a=0.597422, \quad b=-0.000166, \quad c=6.8875285, \quad A_{0}=24.9307.
	\ee
	The cross section is finally generalized for collisions with general targets and general $E_{\ga}$  
	\be \sigma^{UN}(A)=\left. \frac{d\sigma^{UN}}{dt}\right\bracevert_{t=0}(E_{\ga})/\kappa(E_{\ga})\cdot \widetilde{c\si}(E_{\ga})\cdot A\cdot R(A). \ee
	
	\subsection{Spin Density Matrix Elements}
	We can either calculate the Spin Density Matrix Elements (SDMEs) using phenomenological model, \textit{i.e.} Regge Theory, or fit the experimental data with appropriate templates. Generally SDMEs depends on both photon energy $E_{\ga}$ (equivalently c.m. energy $\sqrt{s}$) and the vector-meson's polar angle $\tht^{V}_{cm}$ in the centre-of-mass frame of $\ga N$ (equivalently the momentum transfer $t$).
	
	\subsubsection{SDMEs in Regge Theory}
	\begin{figure}[h]
		\centering
		\includegraphics{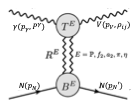}
		\caption{The $V$ photoproduction process in Regge theory}
		\label{fig:diagRT}
	\end{figure}
	At high energies, the vector meson photoproduction process (like $\rho^{0}$) is dominated by Pomeron ($\mathbb{P}$) exchange and natural ($f_{2},a_{2}$) or unnatural ($\pi,\eta$) Regge exchanges subprocesses in Regge theory, as showed in Fig. \ref{fig:diagRT}, although there's large theoretical uncertainty. Fortunately, the J-PARC collaboration performed a fitting to the model parameters using SLAC data and they published the numerical codes for the \textit{linear SDMEs} with their fitted parameters \cite{JPARC}. We extend their numerical calculations to calculate the \textit{circular SDMEs} in Regge theory using the same parameters for our event generator in GEANT4 simulations. The linear SDMEs ($\rho^{0}_{i,j},\rho^{1}_{i,j},\rho^{2}_{i,j}$) take the expressions of Eq.(B1q) to (B1i) in \cite{JPARC}. And the 2 circular SDMEs ($\rho^{3}_{1,0},\rho^{3}_{1,-1}$) take the expression as:
	\be \begin{array}{rcl}
		\Im\rho^{3}_{1,0} &= \frac{\Im}{2N}\sum_{\lam,\lam'}[\mathscr{M}_{\substack{1,1\\ \lam,\lam'}}\mathscr{M}_{\substack{1,0\\ \lam,\lam'}}^{*}-\mathscr{M}_{\substack{-1,1\\ \lam,\lam'}}\mathscr{M}_{\substack{-1,0\\ \lam,\lam'}}^{*}]=& \sum_{\lam,\lam'}\frac{1}{2N}\Im[(\mathscr{M}_{\substack{1,1\\ \lam,\lam'}}+\mathscr{M}_{\substack{1,-1\\ \lam,\lam'}})\mathscr{M}_{\substack{1,0\\ \lam,\lam'}}^{*}]\\
		\Im\rho^{3}_{1,-1} &=  \frac{\Im}{2N}\sum_{\lam,\lam'}[\mathscr{M}_{\substack{1,1\\ \lam,\lam'}}\mathscr{M}_{\substack{1,-1\\ \lam,\lam'}}^{*}-\mathscr{M}_{\substack{-1,1\\ \lam,\lam'}}\mathscr{M}_{\substack{-1,-1\\ \lam,\lam'}}^{*}]=& \sum_{\lam,\lam'}\frac{1}{N}\Im[\mathscr{M}_{\substack{1,1\\ \lam,\lam'}}\mathscr{M}_{\substack{1,-1\\ \lam,\lam'}}^{*}]
	\end{array}\ee
	where the helicity amplitude $\mathscr{M}_{\substack{\lam_{\ga},\lam_{V}\\ \lam,\lam'}}$ depending on $\sqrt{s}$ and $t$, sums over the contributions of Pomeron, natural and unnatural exchanges, and $N$ is the normalized amplitude square when further summing over all helicities.
	\be
	\mathscr{M}_{\substack{\lam_{\ga},\lam_{V}\\ \lam_{N},\lam_{N'}}}(s,t) =\sum_{E=\pi,\eta,\mathbb{P},f_{2},a_{2}} T^{E}_{\lam_{\ga},\lam_{V}}(t)\cdot R^{E}(s,t)\cdot B^{E}_{\lam_{N},\lam_{N'}}(t)
	\ee
	where $T^{E}_{\lam_{\ga},\lam_{V}}(t)$ and $B^{E}_{\lam_{N},\lam_{N'}}(t)$ are the photon vertex and nucleon vertex describing the helicity transfer from the photon ($\ga$) to the vector meson ($V$) and between the nucleon target and recoil, respectively. $R^{E}(s,t)$ is the Regge propagator of the exchanges.
	The general $t-$function form of various SDMEs of a single  $\pi$ or natural exchange can be simplified as:
	\be \label{rhoform}
	\rho^{\alpha}_{i,j}(t)\sim \beta_{0}+\frac{\beta_{1}t+\beta_{2}\sqrt{-t}t+\beta_{3}t^2}{1+\kappa_{1}t+\kappa_{2}t^2}
	\ee
	This form will also be inspiring for our fitting scheme of the SDMEs.
	
	\subsubsection{Linear Spin Density Matrix Elements in Fitting}
	For various Spin Density Matrix Elements $\rho^{\af}_{ij}$, we perform the 2-dimensional $(t,E_{\ga})$ fitting, since $\tht^{V}_{cm}$ depends on $t$ in the inverse function of $t(\tht^{V}_{cm})$. It was found that distributions of SDMEs in the $E_{\ga}$ direction are much smoother than in the $t$ direction \cite{slac} and there is little correlation between  $t-E_{\ga}$ dependences \cite{glueX}, we thus apply a simple strategy that we fit the 3 photon energy measurements of $\left. \rho^{\af}_{ij}(t)\right|_{E_{\ga}}$ along $t$ to get the parameters of the fitting functions $\left. \vec{\bt}^{\af}_{ij}\right|_{E_{\ga}}$. And then fit various $\left. \vec{\bt}^{\af}_{ij}\right|_{E_{\ga}}$ along $E_{\ga}$. With the fitted $\vec{\bt}^{\af}_{ij}(E_{\ga})$ we can build up the 2-dim functions $\rho^{\af}_{ij}(t,E_{\ga})$. All these fittings are performed with ODR method to take the uncertainties into account. 	
	There are 7 $|t|$ ranges of measurements in the first 9 unpolarized and linear polarized SDMEs $\{\rho^{0}_{00},\rho^{0}_{1-1},\Re\rho^{0}_{10},\rho^{1}_{00},\rho^{1}_{11}\rho^{1}_{1-1},\Re\rho^{1}_{10},\Im\rho^{2}_{10},\Im\rho^{2}_{1-1}\}\equiv \dot{\rho}^{k},k=1,\cdots,9$. We use the centre points of the ranges to represent the $|t|$ data: $0.035, 0.065, 0.10, 0.15, 0.215, 0.325, 0.6 (GeV^{2})$.  
	
	To guarantee the SDMEs in physical regions $|\rho^{\af}_{ij}(t)|<1$ even when $|t|$ and $E_{\ga}$ are large, rather than the polynomial trial templates, we use an optimized fitting template inspired by Eq.(\ref{rhoform}):
	\be \begin{split}
		\left. \dot{\rho}^{k}(t)\right|_{E_{\ga}}&=(\left. \bt^{k}_{0}\right|_{E_{\ga}}\sqrt{|t|}|t|+\left. \bt^{k}_{1}\right|_{E_{\ga}}\cdot |t|+\left. \bt^{k}_{2}\right|_{E_{\ga}}\cdot t^{2})/(1+1.6\cdot t^{2}), \\
		\bt^{k}_{l}&(E_{\ga})=(\mathring{\bt}^{k}_{l}\sqrt{E_{\ga}}+\bar{\bt}^{k}_{l}\cdot E_{\ga}+\hat{\bt}^{k}_{l}\cdot E_{\ga}^{2})/[1+1.6(E_{\ga}/E_{0}-E_{0}/E_{\ga})^{2}], \quad l=0,1,2; k=1,\cdots,9
	\end{split} \ee

	\subsubsection{Circular Spin Density Matrix Elements in Fitting} 
	The 2 circular SDMEs $\{\Im\rho^{3}_{10},\Im\rho^{3}_{1-1}\}\equiv\dot{\rho}^{k},k=10,1-1$ in the fitting scheme are more complicated. The SLAC experiment used a linearly polarized photon beam so we need other experiments to extract them. Fortunately, HERMES experiment at DESY has performed the measurements of SDMEs for the virtual photons from electrons $e^{-}\rightarrow \ga^{*}e^{-}$ collide protons or euterons producing $\rho^{0}$ mesons: $\ga^{*}+N\rightarrow \rho^{0}(\rightarrow\pi^{+}\pi^{-})+N'$ \cite{Hermes}.
	They actually tried to measure the scaled SDMEs ratio $r^{\af}_{ij}$ and we focus on the circular polarization $\af=3$.
	\be \label{rhor}
	r^{\af}_{\lam_{V}\lam'_{V}}(Q^{2},t)=\rho^{\af}_{\lam_{V}\lam'_{V}}(Q^{2},t)/[1+\ep(Q^{2}) R(Q^{2})], \quad \af=1,2,3
	\ee
	With the non-zero photon virtuality $Q^{2}$ and $t$ dependence. $R$ is the longitudinal-to-transverse cross section ratio and $\ep$ represents the ratio of fluxes of longitudinal and transverse virtual photons.
	The $R(Q^{2})=c_{0}\left[Q^{2}/m^{2}_{V} \right]^{c_{1}}$ expression is suggested by VMD models and fitted with the HERMES data obtaining $c_{0}=0.56$ and $c_{1}=0.47 $. The $Q^{2}-$binned $\ep$ was measured earlier by HERMES \cite{Hermes2000} and we fit their data to get: $\ep(Q^{2})=a+b Q^{2}+c/(Q^{2}-d)$.
	
	However, HERMES experiment only measured the 1D $t-$binned $\hat{r}^{\af}_{ij}(t)\equiv\langle r^{\af}_{ij}(Q^{2},t) \rangle_{Q^{2}}$ by averaging over $Q^{2}$ and $Q^{2}-$binned $\check{r}^{\af}_{ij}(Q^{2})\equiv\langle r^{\af}_{ij}(Q^{2},t)\rangle_{t}$ by averaging over $t$ \cite{Hermes}. To obtain $\rho^{3}_{ij}$ for real photons we need to solve for the 2D functions $\rho^{3}_{ij}(Q^{2},t)$ from these measurements and extrapolate to the value $\rho^{3}_{ij}(Q^{2}=0,t)$! 
	Unfortunately, the data of $\hat{r}^{\af}_{ij}(t)$ and $\check{r}^{\af}_{ij}(Q^{2})$ are not enough to solve the 2D functions, we have to assume that the $t$ and $Q^{2}$ dependences in $\rho^{3}_{ij}(Q^{2},t)$ are \textbf{separable}: $\rho^{3}_{ij}(Q^{2},t)=\tilde{\rho}^{3}_{ij}(t)\cdot \breve{\rho}^{3}_{ij}(Q^{2})$ and quadratic in $Q^{2}$, which was also implied by the SDMEs measurement of $\omega$ meson \cite{omega}. In this approximation, inserting Eq.(\ref{rhor}) and averaging the obtained $R(Q^{2})\ep(Q^{2})$, we get the solution:
	\be
	\rho^{3}_{ij}(Q^{2}=0,t)=\hat{r}^{3}_{ij}(t)  \left[1 - C_{ij}\delta\tilde{\rho}^{ij}_{1} \left\langle \frac{Q^{2}}{1+\ep R(Q^{2})}\right\rangle_{Q^{2}}  - C_{ij}\delta\tilde{\rho}^{ij}_{2} \left\langle \frac{Q^{4}}{1+\ep R(Q^{2})}\right\rangle_{Q^{2}} \right]  \Bigg/  \left\langle \frac{1}{1+\ep R(Q^{2})}\right\rangle_{Q^{2}} \equiv \hat{r}^{3}_{ij}(t) \cdot \Delta^{3}_{ij}
	\ee
	$C_{ij}$ and $\delta\tilde{\rho}^{ij}_{1,2}$ are the quadratic coefficients determined by $\check{r}^{3}_{ij}(Q^{2})$. The parts except $\hat{r}^{3}_{ij}(t)$ are just constants after $Q^{2}$ average and we denote them together as $\Delta^{3}_{ij}$. We can further calculate a consistency correction for our approximation and fittings, and then apply this correction on the scaling constants $\Delta^{3}_{10}, \Delta^{3}_{1-1}$ by further average $\left\langle \hat{r}^{3}_{ij}(t)\right\rangle_{t}$ and fitting with $\check{r}^{\af}_{ij}(Q^{2})$ data. Thus we fit $\rho^{3}_{ij}(Q^{2}=0,t)$ to the $\hat{r}^{3}_{ij}(t)$ data with the above equation.
	
	We apply an optimized fitting on circular SDMEs inspired by Eg.(\ref{rhoform}): 
	\be
	\dot{\rho}^{k}(t)=\bt^{k}_{0}+\sqrt{|t'|} |t'|/(1+ \bt^{k}_{1}\cdot |t'| + \bt^{k}_{2}\cdot t'^{2}),~ k=10,1-1, \quad t'=t-t_{0}, \quad t_{0}=2E^{cm}_{\ga}(E^{cm}_{V}-P^{cm}_{V})-\mu^{2}
	\ee
	to the $\hat{r}^{3}_{ij}(t)$ data and finally obtain the circular SDMEs $\rho^{3}_{ij}(Q^{2}=0,t)$ which are independent of the photon energy $E_{\ga}$, since the HERMES experiment have no energy variation. The detail and results of the calculations are presented in Appendix (1).
	
	\subsubsection{Discussion on the 2 Schemes} 
\begin{figure}[h]
		\includegraphics{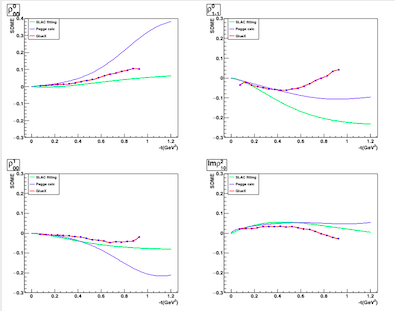}
		\caption{The comparison of SDMEs $\rho_{00}^{0}, \rho_{1-1}^{0}, \rho_{00}^{1}$ and $\rho_{10}^{2}$. }
		\label{fig:SDMEcomp}
	\end{figure}

	Recently the GlueX experiment also presents their preliminary results \cite{glueX} of the linear polarizations of $9 GeV$ photons, and their SDMEs also show a $\Or(t^{2})$ behavior. We compare our SDMEs in both Regge Scheme and Fitting Scheme at $9 GeV$ with their data for $\rho_{00}^{0}, \rho_{1-1}^{0}, \rho_{00}^{1}$ and $\rho_{10}^{2}$ in Fig. (\ref{fig:SDMEcomp}). There are more $t$ points in GlueX SDMEs data but they are just preliminary and have not yet published. Besides, they only have a single photon energy point, so we sticked our fitting on the SLAC data.
	
	On the other hand, the calculations of the circular SDMEs from Regge Scheme showed a wrong sign with the experimental data \cite{Hermes2000}! So we consider the linear Regge Scheme + circular Fitting Scheme or the pure Fitting Scheme be the recommended SDMEs calculations in the GEANT4 simulations.

	\section{Measurement Phenology}
	To measure the 3 polarization components of photons, the angular observables $\theta_{\pi},\phi_{\pi}$ need to be fitted to determine $P^{\ga}_{i}$ from Eq.(\ref{Wfunc}). Observing Eq.(\ref{Wfunc}) we can find it doesn't matter which pion to define $\theta_{\pi},\phi_{\pi}$: when the angles$\theta_{\pi},\phi_{\pi}$ are changed between the definition with respect to $\pi^{+}$ and $\pi^{-}$, 
	\be 
	\theta_{\pi} \rightarrow \pi - \theta_{\pi}, \qquad \text{and} \phi_{\pi} \rightarrow  \pi + \phi_{\pi} 
	\ee
	thus, the terms of $\sin 2\theta_{\pi}$ are unchanged. Therefore, the expressions in Eq.(\ref{Wfunc}) remain the same! From this observation, the \textit{charge identification} in the detectors is not necessary for the photon polarization measurement in our $V$-Photoproduction induced polarimetry! We can actually propose this polarimetry on the space-borne detectors without magnetic field like Fermi-LAT.
	
	\begin{figure}[h] 
		\centering
		\includegraphics{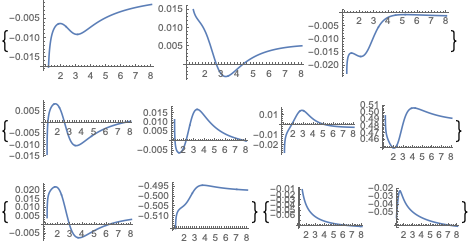}
		\caption{The averages of the 11 SDMEs depending on the photon energy  $E_{\ga} (GeV)$.}
		\label{fig:SDMEsavg}
	\end{figure}
	To measure the 4 components of the photon polarization vector $\vec{P}^{\ga}$, we need to adopt the $\theta_{\pi},\phi_{\pi}$ 2-dimensional fitting or the moment method on the event distributions to evaluate $P^{\ga}_{i}$ using Eq.(\ref{Wfunc}). However, the SDMEs $\rho^{\af}_{ij}(t,E_{\ga})$ depend on squared momentum transfer $t$ or also photon energy $E_{\ga}$. They can be averaged with the $d\si/dt$ weight: $\langle\rho^{\af}_{ij}\rangle$ as functions of  $E_{\ga}$ showed in Fig.(\ref{fig:SDMEsavg}). To simplify the $P^{\ga}_{i}$ measurement and increase the statistics remarkably, we can fit the events with the 4 integrated $\theta_{\pi}$ and $\phi_{\pi}$ $1D$ functions $W_{\theta,\phi}^{\af}$ using these averaged $\langle\rho^{\af}_{ij}\rangle$ and integrating Eq.(\ref{Wfunc}) by $\phi_{\pi}$ and $\theta_{\pi}$. \textit{E.g.}, when $E_{\ga}=3.2GeV$ in Fitting Scheme of SDMEs,
	\be \begin{aligned}
		4W_{\phi}^{0}(\phi_{\pi})&=3.11342 + 0.0391892 \cos 2\phi_{\pi}, & 4W_{\theta}^{0}(\theta_{\pi})&=3.11342 - 3.2261 \cos 2\theta_{\pi}, \\
		4W_{\phi}^{1}(\phi_{\pi})&=0.046692 - 3.11348  \cos 2\phi_{\pi}, & 4W_{\theta}^{1}(\theta_{\pi})&=0.046692 + 0.165025  \cos 2\theta_{\pi}, \\
		4W_{\phi}^{2}(\phi_{\pi})&=-3.10974  \sin 2\phi_{\pi}, & 4W_{\theta}^{2}(\theta_{\pi})&=-0.0517224  \sin 2\theta_{\pi}, \\
		4W_{\phi}^{3}(\phi_{\pi})&=-0.325964  \sin 2\phi_{\pi}, & 4W_{\theta}^{3}(\theta_{\pi})&=-0.733559  \sin 2\theta_{\pi}.
	\end{aligned} \ee
	We observe that the functions $W_{\theta,\phi}^{2}$ and $W_{\theta,\phi}^{3}$ are simply degenerate, $P^{\ga}_{2}$ and $P^{\ga}_{3}$ cannot be separated by fitting $W_{\phi}(\phi_{\pi})$ or $W_{\theta}(\theta_{\pi})$ alone. We propose a \textit{combined-fitting scheme} that we fit $P^{\ga}_{0}$, $P^{\ga}_{1}$ and $P^{\ga}_{2}\langle\rho^{2}_{10}\rangle$ using $W_{\theta}(\theta_{\pi})$ first, and then insert the fitted results to $W_{\phi}(\phi_{\pi})$ and fit for $P^{\ga}_{2}$ and $P^{\ga}_{3}$. In this way the 4 polarization components can be obtained with a $1D$ fitting to reduce the uncertainty.
	
	We implemented our calculations of $V$-Photoproduction induced polarimetry in the Monte Carlo (MC) event generators, \textit{e.g.,} in the detector simulation GEANT4 \cite{polGEANT}. We showed the open angle between $\pi^{+}\pi^{-}$ is really much larger than that between $e^{+}e^{-}$ of high-energy photons from the MC, and thus improve the efficiency of the measurement significantly. We compared the distributions of $\theta_{\pi}$ and $\phi_{\pi}$ of different energy photons in unpolarized, linearly-polarized and circularly-polarized modes from the MC, and showed the validity of our polarimetry! 
	
	\section{Conclusion}	
	We look forward to more high-quality experimental results of both linear and circular SDMEs at more $t$ and $E_{\ga}$ points in the community in the near future. Our model, calculations and fittings will be the accomplished machinery for these new data and will provide a complete polarimetry of high accuracy for high-energy photon! Our new polarimetry induced Vector Meson Photo-production can measure the circular polarization component of high-energy cosmic $\ga$ and the polarization of high-energy cosmic $e^{+}/e^{-}$ for the first time. We calculate the production process of $\pi^{+}\pi^{-}$ by a generally polarized photon near nucleon's field in a generalized \textit{VPD-SDMEs Factorization} with the fitted experimental data, so that it's partially model-independent. We also propose the observables and approach to measure their polarizations based on our calculations.  We look forward to applying our polarimetry in the running, and proposing astroparticle experiments and discovering BSM new physics in high-energy particle processes and revealing the remained mysteries in astrophysical processes in the universe.




\end{document}